\documentclass[aps,pra,twocolumn,superscriptaddress,amsmath,amssymb,tightenlines,epsfig,floatfix]{revtex4}
\usepackage{graphicx}
\usepackage{dcolumn}	
\usepackage{bm}			
\usepackage{amsfonts}
\usepackage{xspace}

\begin{document}

\newcommand{\psihat}{\ensuremath{\hat{\psi}}\xspace}
\newcommand{\psihatd}{\ensuremath{\hat{\psi}^{\dagger}}\xspace}
\newcommand{\ahat}{\ensuremath{\hat{a}}\xspace}
\newcommand{\ahatd}{\ensuremath{\hat{a}^{\dagger}}\xspace}
\newcommand{\bhat}{\ensuremath{\hat{b}}\xspace}
\newcommand{\bhatd}{\ensuremath{\hat{b}^{\dagger}}\xspace}
\newcommand{\boldr}{\ensuremath{\mathbf{r}}\xspace}
\newcommand{\dr}{\ensuremath{\,d^3\mathbf{r}}\xspace}
\newcommand{\etal}{\emph{et al.\/}\xspace}
\newcommand{\ie}{i.e.\:}
\newcommand{\eq}[1]{Eq.\,(\ref{#1})\xspace}
\newcommand{\fig}[1]{Figure\,(\ref{#1})\xspace}
\newcommand{\abs}[1]{\left| #1 \right|}

\title{Self-induced Spatial Dynamics to Enhance Spin Squeezing via One-Axis Twisting in a Two-Component Bose-Einstein Condensate}
\author{S. A. Haine}
\affiliation{University of Queensland, Brisbane, 4072, Australia}
\email{haine@physics.uq.edu.au}
\author{J. Lau}
\affiliation{University of Queensland, Brisbane, 4072, Australia}
\author{R. P. Anderson}
\affiliation{School of Physics, Monash University, Victoria 3800, Australia}
\preprint{Version: submitted \today}

\author{M. T. Johnsson}
\affiliation{Australian National University,  ACT, 0200, Australia}
\preprint{Version: submitted \today}

\begin{abstract}
We theoretically investigate a scheme to enhance relative number squeezing and spin squeezing in a two-component Bose-Einstein condensate (BEC) by utilizing the inherent mean-field dynamics of the condensate.  Due to the asymmetry in the scattering lengths, the two components exhibit large density oscillations where they spatially separate and recombine. The effective non-linearity responsible for the squeezing is increased by up to three orders of magnitude when the two components spatially separate. We perform a multi-mode simulation of the system using the truncated Wigner method, and show that this method can be used to create significant  squeezing in systems where the effective nonlinearity would ordinarily be too small to produce any significant squeezing in sensible time frames, and that strong spatial dynamics resulting from large particle numbers aren't necessarily detrimental to generating squeezing. We develop a simplified semi-analytic model that gives good agreement with our multi-mode simulation, and will be useful for predicting squeezing in a range of different systems. 
\end{abstract}

\maketitle
\section{Introduction}
 In recent years, there has been much interest in atom interferometry for high-precision inertial measurements \cite{chu99, chu2001, kasevich02, kasevich97, kasevichG, altinET2013}, as well as measurements of the fine structure constant \cite{biraben_alpha}, and potentially gravitational wave detection \cite{kasevich_GW}. Although thermal sources of atoms currently have a larger flux, Bose-Einstein condensates have an advantage over thermal atoms as they have a narrower velocity distribution and larger coherence length, allowing for easier manipulation of the motional state and increased visibility \cite{debs2011, szigeti2012}. However, in any interferometer that utilises uncorrelated particles, our ability to estimate an applied phase shift $\phi$ is limited by the standard quantum limit (SQL), $\Delta \phi = 1/\sqrt{N_t}$, where $N_t$ is the total number of detected particles \cite{dowling}.

There has recently been much interest in the use of spin squeezed states of ultra-cold atoms, as it enables atom interferometry with sensitivity beyond the standard quantum limit (SQL) \cite{oberthaler2010, treutlein2010, leroux2010, lucke2011}. Spin squeezing via one-axis twisting \cite{ueda, sorensen2001, sorensen2002} has previously been demonstrated in two-component BECs \cite{oberthaler2010, treutlein2010}. The rate at which spin-squeezing occurs is governed by the parameter $\chi = \chi_{11}+\chi_{22}-2\chi_{12}$, where
\begin{equation}
\label{chiij}
\chi_{ij} = \frac{4 \pi \hbar }{m}\frac{a_{ij}}{N_i N_j} \int n_i(\boldr) n_j(\boldr) \dr \, ,
\end{equation} 
where $m$ is the mass of the atom, $N_i$ and $n_i(\boldr)$ are the population and number density of atoms in component $i$, and $a_{ii}$ and $a_{ij}$ are the inter- and intra-component $s$-wave scattering lengths  \cite{gross2012}. However, in some atomic species, $\chi$ is too small to create significant spin squeezing in any reasonable time. In Rubidium 87 for example, where the relevant atomic states are the $F=1$ and $F=2$ hyperfine ground states, $a_{11}+a_{12}-2 a_{12} \approx 8\times10^{-4} a_{11}$. Despite this, spin squeezing in Rb 87 BECs has been demonstrated by manipulation of one of the scattering lengths via a Feshbach resonance \cite{oberthaler2010} to increase $\chi$. Spin squeezing has also been demonstrated by manipulating the external confining potential of each spin component to separate them spatially, thereby decreasing $\chi_{12}$ in \eq{chiij} and increasing $\chi$ \cite{treutlein2010}. These schemes used BECs containing only a few thousand atoms, as a higher atom number increases the interaction energy, compromising the single-mode behaviour upon which these schemes require. Spin squeezing via one-axis twisting in non-condensed samples of Rb 87 has also been achieved by manufacturing an artificial nonlinearity via coupling to an optical cavity \cite{leroux2010}. 

In this paper we demonstrate a considerably simpler scheme to obtain both relative number and spin squeezing that does not require precise magnetic field control for Feshbach resonances, time- and state-dependent potentials, or optical cavities, considerably simplifying the process. Our scheme utilizes the inherent mean-field dynamics of the two-component system --- which arise from the slight asymmetry in the s-wave scattering lengths and periodically decrease the spatial overlap of the two components --- to create a much higher $\chi$, leading to significant squeezing. Furthermore, it demonstrates that strong multimode dynamics aren't necessarily detrimental to generating spin squeezing, allowing the possibility of spin-squeezing via one-axis twisting in BECs with a large, metrologically useful number of atoms. 
 
\indent The remainder of this paper is organised as follows: In Section II, we describe our spin squeezing scheme and present a multi-mode simulation of the quantum dynamics using the truncated Wigner approach, which has been shown to be highly successful in simulating such systems \cite{steel, sinatra2002, johnsson2007, haine2009, haine2011, Egorov2012, dall2009, dennis2010, johnsson2013, haine2013}. In Section III, we derive an effective two-mode semi-analytic model, and discuss validity of this model. In Section IV, we discuss how the level of squeezing can be controlled by changing the strength of the trapping potential and performing multiple $\pi$ pulses, and the effect of multi-mode dynamics on the mode-overlap.  In Section V, we discuss the usefulness of this scheme for enhanced atom-interferometry. 

\section{Enhancement of Spin squeezing via self-induced dynamics}\label{sec2}
\begin{figure}
\includegraphics[width=1.0\columnwidth]{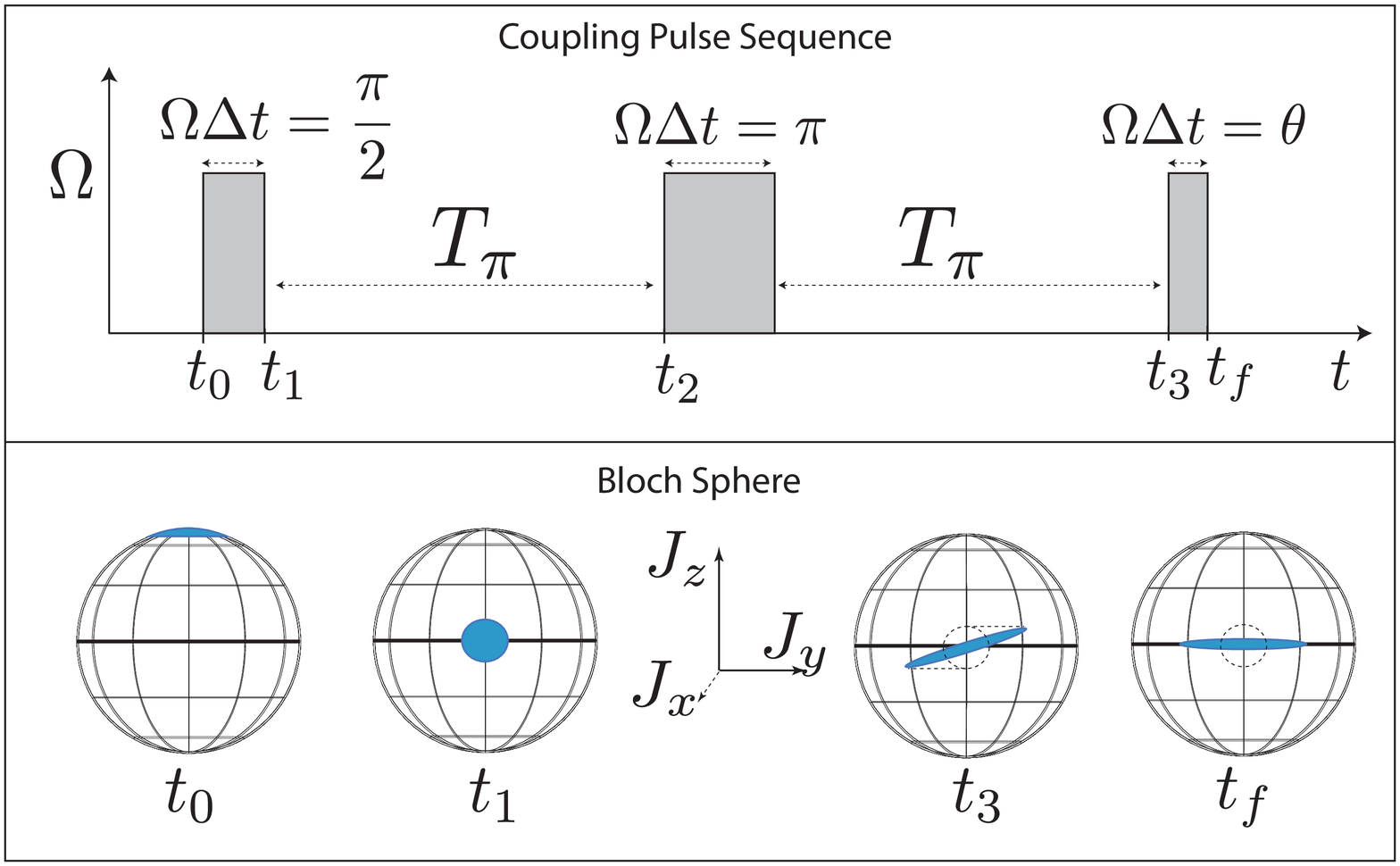}
\caption{Sequence for the coupling pulses used in the scheme, and their effect on the Bloch-sphere.}
\label{fig:scheme}
\end{figure}

Our spin squeezing scheme follows the one-axis twisting scheme  \cite{ueda, oberthaler2010}, for which there has been much theoretical interest \cite{ueda, sorensen2001, sorensen2002, Li2008, sinatrareview2012, ferrini2011}. Our scheme is outlined in \fig{fig:scheme}. We consider a $^{87}$Rb BEC with two hyperfine levels, $|a\rangle \equiv |F=1,m=-1\rangle$ and $|b\rangle \equiv |F=2,m=+1\rangle$ confined in a spherically symmetric harmonic potential. In this proposal, all of the condensate atoms are initially prepared in the $|a\rangle$ state and then apply a short $\pi/2$-microwave coupling pulse to transfer half of the population of the atoms into the $|b\rangle$ state. The system is then left to evolve for a period of free evolution, in the absence of any microwave coupling. During this period, nonlinear interactions between the atoms and the slight asymmetry in the scattering lengths then causes the wave function of the two components to spatially separate and recombine in an oscillatory manner \cite{anderson2009,egorov2011}. The parameter governing the squeezing rate, $\chi$ is greatly increased when the two components spatially separate \cite{treutlein2010}. A spin echo pulse is applied at the midpoint of the free evolution to correct for dephasing effects due to uncertainty in the total number of particles \cite{atlin}. In particular parameter regimes, the wave functions of the two components approximately overlap at the end of the free evolution period, allowing for high-contrast interferometry between the two modes. A second microwave coupling pulse is then applied for a variable time $t_\theta$. The behaviour of the system is intuitive in the context of the three pseudo spin operators
\begin{eqnarray}
\label{psuedospin}
\hat{J}_x &=& \frac{1}{2}\int \left( \psihatd_b(\boldr) \psihat_a(\boldr) + \psihatd_a(\boldr) \psihat_b(\boldr)\right) \dr \\
\hat{J}_y &=& \frac{i}{2}\int \left( \psihatd_b(\boldr) \psihat_a(\boldr) - \psihatd_a(\boldr) \psihat_b(\boldr)\right) \dr \\
\hat{J}_z &=& \frac{1}{2}\int \left( \psihatd_a(\boldr) \psihat_a(\boldr) - \psihatd_b(\boldr) \psihat_b(\boldr)\right) \dr \\
&=& (\hat{N}_a - \hat{N}_b) / 2
\end{eqnarray}
where $\psihat_{i}(\boldr)$ is the annihilation operator for an atom at position $\boldr$ in hyperfine state $|i\rangle$, and 
\begin{equation}
\hat{N}_j = \int \psihatd_j(\boldr)\psihat_j(\boldr) \, \dr \, 
\end{equation}
 is the number operator for atoms in hyperfine state $|i\rangle$, where $i = a, b$. We begin with the spin expectation value at the north pole of the Bloch sphere (\fig{fig:scheme}). The first coupling pulse rotates the spin expectation value to the equator. During the period of free evolution, inter-particle interactions cause a nonlinear phase shift, shearing the uncertainty of condensate spin, as well as a drift around the equator. A $\pi$ pulse followed by another period of free evolution reverses the effect of the drift, while maintaining the shearing. At the end of the free evolution a phase shift of $\frac{\pi}{2}$ rotates the state to lie along the $J_x$ axis, where the final adjustable coupling pulse rotates the state by an amount $\theta = \Omega_0 t_\theta$. This rotation angle is required to rotate the squeezed quadrature into the $J_z$ basis such that it can be directly detected by measuring the population difference between the two components. Unlike the two previous experimental schemes, our scheme does not rely on using Feshbach resonance \cite{oberthaler2010} or a state dependent potential \cite{treutlein2010}  to enhance the effective non-linearity of the two component rubidium BEC. Instead, we utilize the inherent mean-field dynamics of the two components to enhance squeezing. This requires only adjustment of the trap frequencies and timing of coupling pulses.\\

Assuming the microwave field is on resonance for the $|a\rangle \rightarrow |b\rangle$ transition, and making the rotating wave approximation, the effective many body Hamiltonian which describes the quantum dynamics of the two component condensate is given by $\hat{\mathcal{H}}=\hat{\mathcal{H}}_0+\hat{\mathcal{H}}_c$ \cite{haine2009} where
\begin{eqnarray}  
\hat{\mathcal{H}}_0 &=& \sum_{i=a,b}\int \hat{\psi}_i^\dag(\textbf{r})\hat{H}_i\hat{\psi}_i(\textbf{r}) \ensuremath{\,d^3\mathbf{r}} \nonumber \\
&+& \sum_{i,j=a,b}\frac{U_{ij}}{2} \int \hat{\psi}_i^\dag (\textbf{r})\hat{\psi}_j^\dag(\textbf{r}) \hat{\psi}_i(\textbf{r}) \hat{\psi}_j(\textbf{r}) \ensuremath{\,d^3\mathbf{r}}, \label{H0} 
\end{eqnarray}
and
\begin{equation}
\hat{\mathcal{H}}_c(t)=\int \left(\hbar\frac{\Omega(t)}{2} \hat{\psi}_a^\dag(\textbf{r})\hat{\psi}_b(\textbf{r}) e^{i\delta t}+ h.c.\right) \, \dr. \label{Hc}
\end{equation}
$\hat{\mathcal{H}}_0$ describes the the free evolution of the two component BEC, whereas $\hat{\mathcal{H}}_c$ describes the microwave coupling field which is only present when the coupling field is applied.   $\hat{H}_a =  \hat{H}_0\equiv \frac{-\hbar^2}{2m}\nabla^2+V(\textbf{r})$ and $\hat{H}_b = \hat{H}_0 + \hbar \delta$ represent the single particle Hamiltonian where $\hbar \delta$ is the energy difference between the hyperfine states $|a\rangle$ and $|b\rangle$ and $V(\boldr) = \frac{1}{2}m\omega_r^2 r^2$ is the trapping potential. $U_{ij}$ is the nonlinear interaction potential and is given by $U_{ij}=4\pi\hbar^2a_{ij}/m$ where $a_{ij}$ is the $s$-wave scattering length between $|i\rangle$ and $|j\rangle$. The scattering lengths for a two component $^{87}$Rb condensate are taken to be $a_{11} = 100.4 a_0 $, $a_{22} = 95.00 a_0$ and $a_{12} = 97.66 a_0$ \cite{mertes2007}.  $\Omega(t)=\Omega_0 f(t)e^{i\phi}$ represents the coupling field where $\Omega_0$ is the Rabi frequency, $f(t)$ is a function that can be switched between $0$ and $1$ to turn the coupling on and off, and $\phi$ is the phase of the microwave field. Adjusting $\phi$ during the final coupling pulse is equivalent to altering the relative phase of the two atomic wave functions. By making the transformation $\psihat_b \rightarrow \psihat_b e^{i\delta t}$, the Heisenberg equations of motion become

\begin{eqnarray} \label{Heisenberg}
i\hbar\frac{\partial \hat{\psi}_a(\textbf{r})}{\partial t} &= \hat{\mathcal{L}}_a\psihat_a(\boldr) +  \frac{1}{2} \hbar \Omega(t) \hat{\psi}_b(\textbf{r}),\\
i\hbar\frac{\partial \hat{\psi}_b(\textbf{r})}{\partial t} &= \hat{\mathcal{L}}_b\psihat_b(\boldr) + \frac{1}{2} \hbar \Omega^*(t) \hat{\psi}_a(\textbf{r}) ,
\end{eqnarray}
where 
\begin{eqnarray}
\hat{\mathcal{L}}_a &=&\hat{H}_0 + U_{aa}\psihatd_a\psihat_a + U_{ab}\psihatd_b\psihat_b  \\
\hat{\mathcal{L}}_b &=&\hat{H}_0 + U_{ab}\psihatd_a\psihat_a + U_{bb}\psihatd_b\psihat_b.
\end{eqnarray}

If we assume that the dynamics of the coupling is fast compared to the dynamics due to the potential, kinetic, and nonlinear terms, it is sufficient to solve for the dynamics of $\frac{\pi}{2}$, $\pi$, and $\theta$ pulses by ignoring the contribution from $\hat{\mathcal{L}}_j$, in which case
\begin{eqnarray}
\psihat_a(\boldr, t_1) &=& \cos \frac{\theta}{2} \, \psihat_a(\boldr, t_0) -i\sin \frac{\theta}{2} \, \psihat_b(\boldr, t_0) e^{i\phi} \\
\psihat_b(\boldr, t_1) &=& \cos \frac{\theta}{2} \, \psihat_b(\boldr, t_0) -i\sin\frac{\theta}{2} \, \psihat_a(\boldr, t_0) e^{-i\phi},
\end{eqnarray}
where $\theta \equiv \Omega_0 (t_1-t_0)$.

To numerically simulate the quantum dynamics of the system during the free evolution period, we proceed by using Truncated Wigner (TW) approximation. Following standard methods \cite{blackie2008, mtj2007}, the Heisenberg equations can be converted into Fokker-Plank equations (FPEs) by using the correspondences between the quantum operators and the Wigner function. By truncating third and higher order terms, the FPEs can be mapped onto a set of stochastic partial differential equations for complex valued fields $\psi_i(\boldr,t)$,  which are very similar to the usual coupled Gross Pitaevskii equations (GPEs). By averaging over many trajectories with different initial conditions, expectation values of quantities corresponding to operators in the full quantum theory can be obtained. Specifically, 
\begin{equation}
\left \langle \left\{ f\left( \psihatd_j(\boldr), \psihat_j(\boldr) \right) \right\}_{\mbox{sym}}  \right \rangle = \overline{f\left(\psi^*_j(\boldr), \psi_j(\boldr)\right)}
\end{equation}
where ``sym'' denotes symmetric ordering \cite{walls}, and the overline denotes the mean over many stochastic trajectories. The initial conditions are sampled from the appropriate Wigner distribution \cite{olsenwig}.

 The equations governing the evolution of the complex fields are

\begin{eqnarray} \label{GPE}
i\hbar\frac{\partial \psi_a(\boldr)}{\partial t} &= \mathcal{L}_a \psi_a(\boldr)+ \frac{1}{2} \hbar \Omega(t) \psi_b(\boldr) \, , \label{stoch_eom1} \\
i\hbar\frac{\partial \psi_b(\boldr)}{\partial t} &= \mathcal{L}_b \psi_b(\boldr)+ \frac{1}{2} \hbar \Omega^*(t) \psi_a(\boldr), \label{stoch_eom2}
\end{eqnarray}
where
\begin{align}
\mathcal{L}_i &= \frac{-\hbar^2}{2m}\nabla^2 + V(\boldr)+U_{ii} \left(|\psi_i(\boldr)|^2-\frac{1}{\Delta v}\right) +  \nonumber \\
&\qquad  U_{ij} \left( |\psi_j(\boldr)|^2 - \frac{1}{2\Delta v} \right), 
\end{align}
where $\Delta v$ is the volume element that characterises the numeric discretisation of the grid. 

\begin{figure}
\includegraphics[width=1.0\columnwidth]{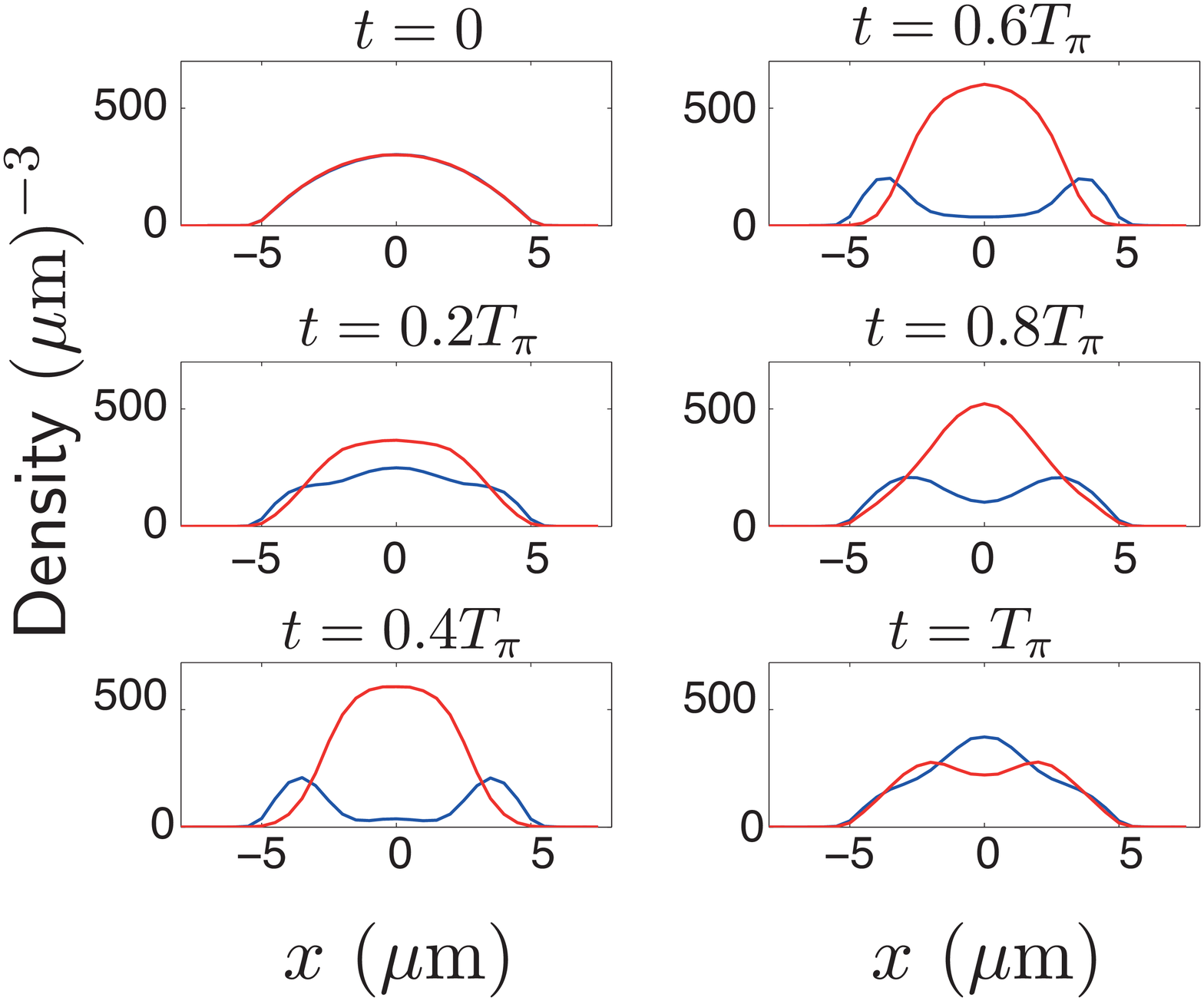}
\caption{Evolution of the density profile after the initial $\pi/2$ coupling pulse equally populates the two components. A slice of the expectation value of the density  $\langle \psihatd_j(\boldr)\psihat_j(\boldr)\rangle$ for each component ($j=a$ (blue), $j=b$ (red)) at $y=z=0$ is shown for several different times. A $\pi$ pulse is applied at $t =T_\pi = 13.29$ ms.  }
\label{densmovie}
\end{figure}

For the purposes of spin squeezing, the behaviour of the system is largely insensitive to the number statistics of the initial state \cite{haine2009}, so for simplicity, we chose our initial state as a Glauber coherent state \cite{walls}. It was shown in \cite{haine2009} that a mixture of coherent states with random phases, or equivalently, a Poissonian mixture of number states, behaves identically to a pure coherent state in this situation. Specifically, we chose the initial state of the system to be $\mathcal{D}(\alpha)|0\rangle$, with
\begin{equation}
\mathcal{D}(\alpha) = \exp{\left(\alpha \ahatd_{g} - \alpha^* \ahat_{g}\right)}\, ,
\end{equation}
with
\begin{equation}
\ahat_{g} = \int_{\mathrm{all space}} \psi^*_g(\boldr) \psihat_a(\boldr)\, \dr \, ,
\end{equation}
where $\psi_g(\boldr)$ is the (normalised) ground state of the Gross-Pitaevskii equation with all the population in $|a\rangle$. The initial conditions in the stochastic simulation that correspond to this situation are
\begin{eqnarray}
\psi_a(\boldr) &=& \sqrt{N_t}\psi_g(\boldr) + \frac{\eta_a(\boldr)}{\sqrt{\Delta v}} \\
\psi_b(\boldr) &=& \frac{\eta_b(\boldr)}{\sqrt{\Delta v}}
\end{eqnarray}
where $N_t = |\alpha|^2$ is the expectation value of the total number of atoms, and $\eta_m(\boldr)$ are complex Gaussian noise functions satisfying $\overline{\eta^*_m(\boldr_i)\eta_n(\boldr_j)} = \frac{1}{2}\delta_{m,n}\delta_{i,j}$. We numerically integrated equations (\ref{stoch_eom1}) and (\ref{stoch_eom2}) using a $32\times32\times32$ spatial grid and $1000$ stochastic trajectories using the XMDS2 numerical integration package \cite{xmds}. The total number of atoms was $1.5\times 10^5$, and the trapping potential was chosen to be a spherically symmetric harmonic potential with radial trapping frequency to be $\omega_r = 200$ rad/s.  Figure~\ref{densmovie} shows cross-section ($y=z=0$) of the expectation value of the density for each component $\langle \psihatd_j(\boldr)\psihat_j(\boldr)\rangle$ for several different times. The two components initially separate, but eventually wobble back together in a quasi-periodic fashion. 

\begin{figure}
\includegraphics[width=0.8\columnwidth]{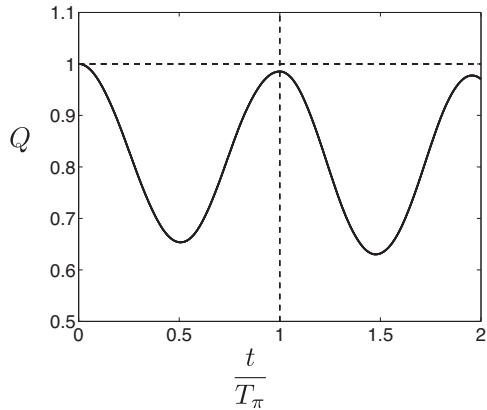}
\caption{Spatial overlap $Q$ of the two components as a function of free-evolution time. At $t=T_\pi$, $Q=0.985$, and at $t=2T_\pi$, $Q= 0.971$. }
\label{fig:overlap}
\end{figure}

The degree to which the two components separate is relevant for enhancing the effective squeezing rate $\chi$. However, in order to convert the spin squeezing along an arbitrary axis to squeezing in $J_z$  (that is, number difference squeezing) that can be directly measured, operations with beam splitters must be performed, which requires good mode-matching, or in other words, a high degree of spatial overlap in the density and phase of the two components. We quantify the overlap as 
\begin{equation}
Q = \frac{1}{\sqrt{\langle \hat{N}_a\rangle \langle \hat{N}_b\rangle}} \abs{ \int  \langle \psihatd_a(\boldr)\psihat_b(\boldr) \rangle  \, \dr } \, ,
\end{equation}
The overlap also has implications for interferometry, as it is directly proportional to the visibility of the fringes. $Q$ is also related to the expectation value of the transverse spin vector $J_{\perp} = \sqrt{\langle \hat{J}_x \rangle^2 + \langle \hat{J}_y \rangle^2} = \sqrt{\langle \hat{N}_1\rangle \langle \hat{N}_2\rangle}\ Q$. \fig{fig:overlap} shows the overlap function $Q$ over time for this system. The two components separate and recombine in a quasi-periodic fashion, with a slight degradation in overlap with each `bounce'. To implement one-axis twisting in this set up, a $\pi$ pulse is implemented at the first revival in overlap $t =T_\pi =  13.29\,$ms, and then the variable angle beamsplitter is implemented at $t = 2T_\pi$. It should be noted that this isn't quite commensurate with maximum overlap, but we chose to keep the total time of free evolution as $t=2 T_\pi$ to minimise phase diffusion from fluctuations in the total number. 

\begin{figure}
\includegraphics[width=1.0\columnwidth]{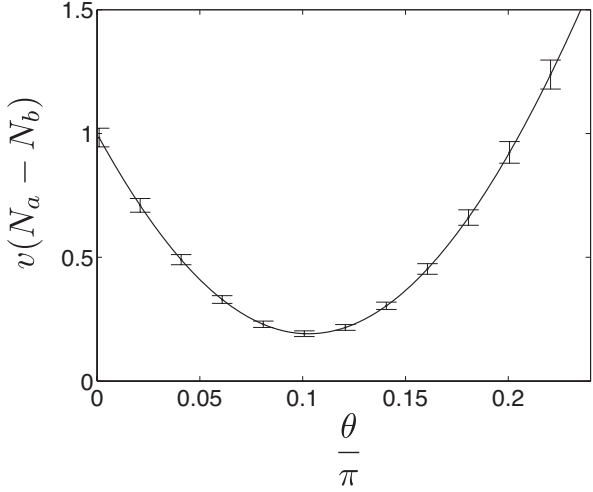}
\caption{$v(N_a-N_b)$ versus final beam-splitter rotation angle $\theta$ at $t=2T_\pi$. The minimum value of $v(N_a-N_b)$ is slightly less than $0.2$ at $\theta = 0.1 \pi$. The error bars are due to stochastic sampling error.  }
\label{fig:var1}
\end{figure}

A state with relative number squeezing is prepared by applying a phase shift of $\pi/2$ before applying a coupling pulse of adjustable angle $\theta$ at $t=2T_{\pi}$. This rotates the squeezing such that the minimum variance is in the $J_z$ direction.  We quantify the squeezing by the normalised variance in the number difference, as this is straightforward to measure directly. We define the normalised number difference variance as
\begin{equation}
v(N_a-N_b) = \frac{\left\langle \left(\hat{N}_a-\hat{N}_b\right)^2 \right\rangle - \left\langle \left(\hat{N}_a-\hat{N}_b \right) \right\rangle^2}{\left\langle \hat{N}_a +  \hat{N}_b \right\rangle} \, ,
\end{equation}
A normalised variance in the number difference of $v(N_a-N_b) <1$ indicates squeezing; $v(N_a-N_b) =1$ is the quantum (or shot-noise) limit, which is the value obtained by dividing a condensate into two equal populations via a linear coupling operation \cite{wineland94}. Alternatively the squeezing could be quantified by the Wineland spin-squeezing parameter \cite{wineland94}
\begin{equation}
\xi_s = \frac{\sqrt{N_t \langle \hat{J}_z^2\rangle}}{J_\perp} = \frac{\sqrt{v(N_a-N_b)}}{Q}
\end{equation}
which is the relevant parameter for enhancing interferometric sensitivity, which is discussed in Section~\ref{sec4}. 
Figure~\ref{fig:var1} shows $v(N_a-N_b)$ vs $\theta$. $v(N_a-N_b)$ dips significantly below $1.0$, indicating significant squeezing can be achieved via this method. 

\section{Semi-analytic model}
As the full 3D TW simulations are very computationally demanding, it is useful to be able to develop a simplified model. We will first develop an analytic two-mode model, which requires only a few input parameters, such as the total number of particles and the effective `squeezing parameter'. We will then develop a model based on the two-component Gross-Pitaevskii \cite{ho1996} equation to estimate the appropriate squeezing parameter, which is used as an input to the two-mode analytic model in order to predict the level of squeezing present in the full multi-mode system. We begin by expanding our field operators over a complete set of time-dependent spatial mode functions 
\begin{eqnarray}
\psihat_a(\boldr,t) &=& \sum_{j} \ahat_j u_{a,j}(\boldr,t) \approx \ahat u_a(\boldr,t) \\
\psihat_b(\boldr,t) &=& \sum_{j} \bhat_j u_{b,j}(\boldr,t) \approx \bhat u_b(\boldr,t) \, .
\end{eqnarray}
where $\ahat\equiv \ahat_0$, $\bhat \equiv \bhat_0$, $u_{a}(\boldr)\equiv u_{a,0}(\boldr)$ and $u_{b}(\boldr)\equiv u_{b,0}(\boldr)$. We have made the approximation that only one mode is significantly occupied. Using this expansion in \eq{H0} and \eq{Hc}, the Hamiltonian becomes
\begin{eqnarray}
\hat{\mathcal{H}} &=& \hbar\chi_{aa}(t) \ahatd\ahatd\ahat\ahat + \hbar\chi_{bb}(t)\bhatd\bhatd\bhat\bhat + 2\chi_{ab}(t)\ahatd\ahat\bhatd\bhat  \nonumber \\
&+& \hbar \delta \bhatd\bhat + \hbar \left(\frac{\Omega(t)}{2} \ahat \bhatd e^{-i\delta t} + \mbox{h.c.}\right),  
\end{eqnarray}
where 
\begin{equation}
\chi_{ij}(t) = \frac{U_{ij}}{2\hbar}\int |u_i(\boldr,t)|^2 |u_j(\boldr,t)|^2 \, \dr \, , \label{chi_ij}
\end{equation}
and we have assumed that, at the times when the coupling is active, $\int u_a^*(\boldr) u_b(\boldr) \dr \approx 1$, which is equivalent to the condition $Q \approx 1$. By transforming to the interaction picture $\bhat \rightarrow \bhat e^{i\delta t}$, we obtain
\begin{eqnarray}
\hat{\mathcal{H}} &=& \hbar\chi_{aa}(t) \ahatd\ahatd\ahat\ahat + \hbar\chi_{bb}(t)\bhatd\bhatd\bhat\bhat + 2\chi_{ab}(t)\ahatd\ahat\bhatd\bhat  \nonumber \\
&+& \hbar \left(\frac{\Omega(t)}{2} \ahat \bhatd + \mbox{h.c.}\right)  .  
\end{eqnarray}
Following the procedure presented in \cite{haine2009}, we choose our initial state to be a Glauber coherent state
\begin{equation}
|\Psi(0)\rangle = |\alpha_0, 0\rangle \, .
\end{equation}
Assuming that the dynamics induced by $\mathcal{H}_c$ occur on a time scale much shorter than the dynamics induced by $\mathcal{H}_0$, after applying a  $\pi/2$ coupling pulse ($\Omega_0 (t_1-t_0) = \pi/2$), we obtain
\begin{equation}
|\Psi(t_1)\rangle = |\alpha(t_1), \beta(t_1)\rangle , 
\end{equation}
with $\alpha(t_1) = \alpha_0/\sqrt{2}$, $\beta(t_1) = -i\alpha_0/\sqrt{2}$.
Expressed in the number basis, this is
\begin{equation}
|\Psi(t_1)\rangle = \sum_{n_1=0}^{\infty}\sum_{n_2=0}^{\infty} C_{n_1,n_2}|n_1, n_2\rangle ,
\end{equation}
with
\begin{equation}
C_{n_1, n_2} = e^{-\frac{1}{2}\left(|\alpha|^2+|\beta|^2\right)}\frac{\alpha(t_1)^{n_1}}{\sqrt{n_1!}} \frac{\beta(t_1)^{n_2}}{\sqrt{n_2!}} .
\end{equation}
During the period of free evolution ($\Omega(t)=0$), the Hamiltonian is diagonal in the number basis, so it is trivial to calculate the evolution of the state. At time $t_2$, after a period $T$ of free evolution, we obtain
\begin{equation}
|\Psi(t_2)\rangle = \sum_{n_1=0}^{\infty}\sum_{n_2=0}^{\infty}C_{n_1, n_2}|n_1, n_2\rangle e^{-i\Phi_{T_1,n_1,n_2}} ,
\end{equation}
with 
\begin{eqnarray}
\Phi_{T_1,n_1, n_2} &=& \int_{0}^{T} \left(\chi_{aa}(t)n_1(n_1-1) + \chi_{bb}(t)n_2(n_2-1) \right.  \nonumber \\ 
 &+& \left. \chi_{ab}(t)n_1n_2\right) dt.   \label{kerrphase1}
\end{eqnarray}
At $t=t_2$ we apply a $\pi$ coupling pulse, which completely exchanges the population between $a$ and $b$. After evolving for another period of time $T$, our final state is
\begin{equation}
|\Psi(t_3)\rangle =\sum_{n_1=0}^{\infty}\sum_{n_2=0}^{\infty}C_{n_1, n_2}|n_1, n_2\rangle e^{-i\Phi_{n_1,n_2}} \, , \label{kerrstate0}
\end{equation}
where $\Phi_{n_1, n_2} = \Phi_{T_1, n_1, n_2}+ \Phi_{T_2, n_1, n_2}$, and 
\begin{eqnarray}
\Phi_{T_2, n_1, n_2} &=& \int_{t_2}^{t_2+T} \left(\chi_{aa}(t)n_2(n_2-1) + \chi_{bb}(t)n_1(n_1-1) \right.  \nonumber \\ 
 &+& \left. \chi_{ab}(t)n_1n_2\right)dt.  \label{kerrstate}
\end{eqnarray}
The evolution due to the final beam splitter is calculated in the Heisenberg picture. Again, by assuming that the contribution due to $\mathcal{H}_0$ is negligible in this time, we obtain
\begin{eqnarray}
\ahat(t_f) &=& \cos \frac{\theta}{2} \ahat(0) -ie^{i\phi}\sin \frac{\theta}{2} \bhat(0) \\
\bhat(t_f) &=& \cos \frac{\theta}{2} \bhat(0) -ie^{-i\phi}\sin\frac{\theta}{2} \ahat(0) \, .
\end{eqnarray}
The number difference becomes
\begin{eqnarray}
\hat{N}_a -\hat{N}_b &=& \ahatd(t_f)\ahat(t_f) -  \bhatd(t_f)\bhat(t_f) \nonumber \\
&=& \cos\theta \left(\ahatd(0)\ahat(0) - \bhatd(0)\bhat(0)\right)  \nonumber \\
&+& i\sin\theta\left(\ahat(0)\bhatd(0) e^{-i\phi} - \bhat(0)\ahatd(0) e^{i\phi}\right) \nonumber \\ \label{vnumdiffop}
\end{eqnarray}
We can calculate the variance in this quantity by calculating the expectation value of the various operator-valued terms in \eq{vnumdiffop} with respect to \eq{kerrstate0}. For example
\begin{eqnarray}
&\,& \langle \Psi(t_3)| \ahatd(0)\bhat(0)|\Psi(t_3)\rangle   \nonumber \\
 &=& \sum_{m_1=0}^{\infty}\sum_{m_2=0}^{\infty} \sum_{n_1=0}^{\infty}\sum_{n_2=1}^{\infty} C_{m_1, m_2}^*C_{n_1, n_2} e^{i\left(\Phi_{m_1, m_2}-\Phi_{n_1, n_2}\right)} \nonumber \\
 &\times & \sqrt{n_1+1}\sqrt{n_2}\langle m_1, m_2|n_1+1, n_2-1\rangle  \nonumber \\
 &=& \sum_{n_1=0}^{\infty}\sum_{n_2=1}^{\infty} \sqrt{n_1+1}\sqrt{n_2}C_{n_1+1, n_2-1}^*C_{n_1, n_2} \nonumber \\
 &\times &e^{i\left(\Phi_{n_1+1, n_2-1}-\Phi_{n_1, n_2}\right)}  \nonumber \\
 &=& \sum_{n_1=0}^{\infty}\sum_{n_2=1}^{\infty} \frac{\alpha^{* n_1+1}}{\sqrt{(n_1+1)!}}\frac{\beta^{*n_2-1}}{\sqrt{(n_2-1)!}}\frac{\alpha^{n_1}}{\sqrt{n_1!}}\frac{\beta^{n_2}}{\sqrt{n_2!}} \nonumber \\
 &\times &  \sqrt{n_1+1}\sqrt{n_2} e^{i 2\left(\lambda_1n_1 - \lambda_2(n_2-1)\right)} e^{-\left(|\alpha|^2 + |\beta|^2\right)}  \nonumber \\
 &=& \sum_{n_1=0}^{\infty}\sum_{n_2=1}^{\infty} \alpha^*\beta \frac{\left(|\alpha|^2 e^{i 2\lambda_1}\right)^{n_1}}{n_1!}\frac{\left(|\beta|^2 e^{-i 2\lambda_2}\right)^{n_2-1}}{(n_2-1)!} \nonumber \\ &\times & e^{-\left(|\alpha|^2 + |\beta|^2\right)} \nonumber \\
 &=&\alpha^*\beta  \exp\left[ |\alpha|^2\left(e^{i2\lambda_1}-1\right)+|\beta|^2\left(e^{-i2\lambda_2}-1\right)\right]
 \end{eqnarray}
where
\begin{eqnarray}
\lambda_1 &=& \int_0^T \left(\chi_{11}(t) -\chi_{12}(t)\right)dt  \nonumber  \\
&+& \int_{t_2}^{t_2+T} \left(\chi_{22}(t) -\chi_{12}(t)\right)dt, \label{r1_def} \\
\lambda_2 &=& \int_0^T \left(\chi_{22}(t) -\chi_{12}(t)\right)dt \nonumber \\
&+& \int_{t_2}^{t_2+T} \left(\chi_{11}(t) -\chi_{12}(t)\right)dt. \label{r2_def}
\end{eqnarray}
If the dynamics in the trap is approximately periodic, then $\int_{0}^T \chi_{ij}(t) dt \approx \int_{t_2}^{t_2+T} \chi_{ij}(t) dt$, in which case $\lambda_1 \approx \lambda_2 \equiv \lambda $, and the relevant parameter that governs the degree of squeezing is 
\begin{equation}
\lambda = \int_{0}^T \chi(t) \, dt , \label{rchi}
\end{equation}
where $\chi(t) = \chi_{11}(t)+\chi_{22}(t)-2\chi_{12}(t)$ is the familiar one-axis twisting rate \cite{gross2012}. A list of operator expectation values required to calculate $v(N_1-N_2)$ with respect to \eq{kerrstate0} is given in Table \ref{kerr_expect_table}. 

 \begin{table}
   \caption{\label{kerr_expect_table} Expectation value of various operators with respect to \eq{kerrstate0}. }
     \begin{ruledtabular}
     \begin{tabular}{ c | c }
    $\hat{X}$ & $\langle \hat{X} \rangle$ \\
    \hline
    $\ahatd\ahat$ & $|\alpha|^2$ \\
    $\bhatd\bhat$ & $|\beta|^2$ \\
    $\ahatd\bhat$ & $\alpha^*\beta  \exp\left[ |\alpha|^2\left(e^{2i\lambda}-1\right)+|\beta|^2\left(e^{-2i\lambda}-1\right)\right]$ \\
    $\ahatd\ahat\bhatd\bhat$ & $|\alpha|^2|\beta|^2$ \\
    $\ahatd\ahat\ahatd\ahat$ & $|\alpha|^4 +|\alpha|^2$ \\
    $\bhatd\bhat\bhatd\bhat$ & $|\beta|^4 +|\beta|^2$ \\
    $\ahatd\ahat\ahat\bhatd$ & $\alpha\beta^* |\alpha|^2 e^{-2i\lambda}\exp\left[ |\alpha|^2\left(e^{-2i\lambda}-1\right)+|\beta|^2\left(e^{2i\lambda}-1\right)\right] $ \\
    $\ahat\bhatd\bhatd\bhat$ & $\alpha\beta^* |\beta|^2 e^{2i\lambda}\exp\left[ |\alpha|^2\left(e^{-2i\lambda}-1\right)+|\beta|^2\left(e^{2i\lambda}-1\right)\right] $ \\
    $\ahatd\ahatd\bhat\bhat$ & $\alpha^{*2}\beta^2 e^{2i\lambda}\exp\left[ |\alpha|^2\left(e^{4i\lambda}-1\right)+|\beta|^2\left(e^{-4i\lambda}-1\right)\right] $ \\
  \end{tabular}
       \end{ruledtabular}
 \end{table}

\begin{figure}
\includegraphics[width=1.0\columnwidth]{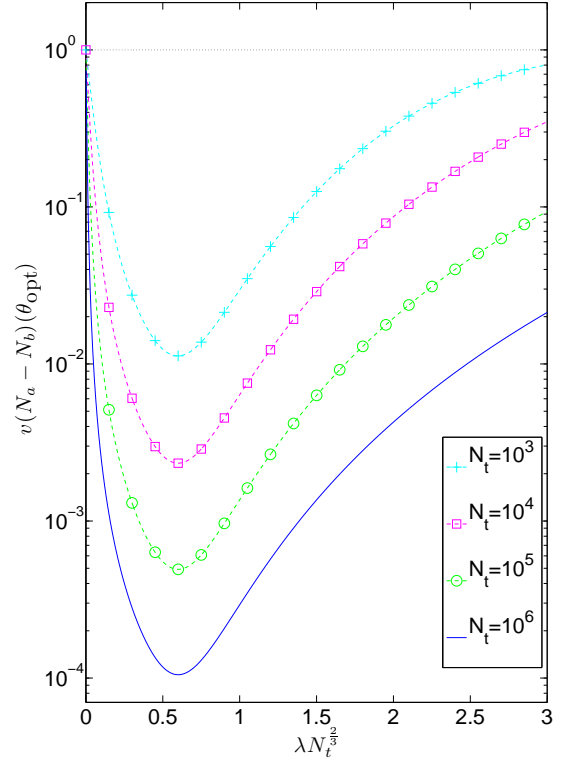}
\caption{\label{fig:var_vs_r} $v(N_a-N_b)$ at $\theta_{\mathrm{opt}}$ vs $\lambda$ for several different values of $N_t$.}
\end{figure}

\begin{figure}
\includegraphics[width=0.9\columnwidth]{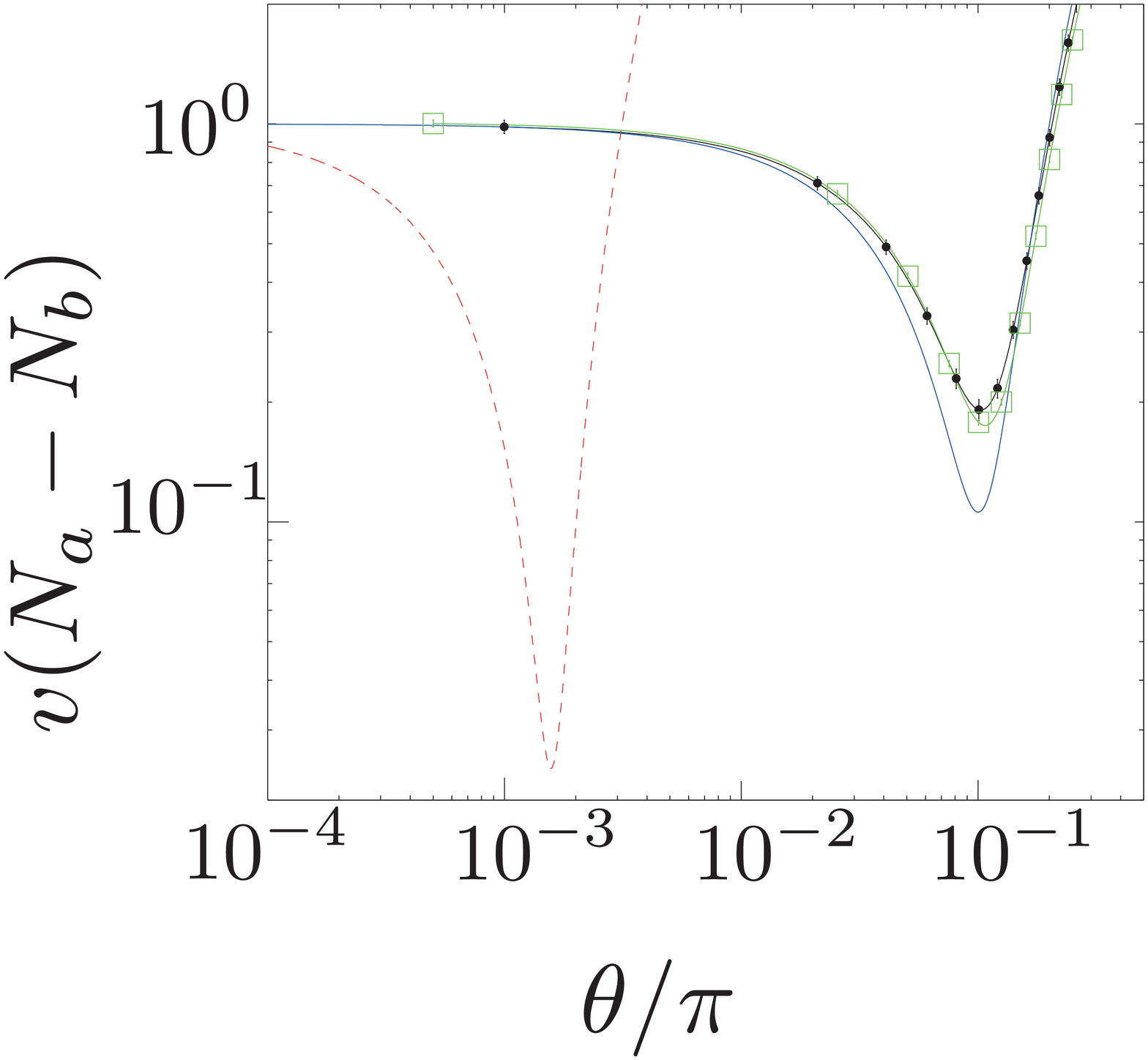}
\caption{\label{fig:v_vs_vanal} Comparison of two-mode model (red dashed line) with full 3D TW model (black dots). The effective squeezing parameter $\lambda = 7.99\times 10^{-4}$ was determined from \eq{rchi}. Much better agreement is given by using $\lambda = 9.18\times 10^{-6}$ calculated from \eq{rphi} (blue solid line). The green squares are the result of a 1D TW simulation with spherical symmetry. The error bars from the 1D simulation are too small to see on this scale.}
\end{figure}

Using the expressions in Table \ref{kerr_expect_table} and their Hermitian conjugates, at $\phi = \pi/2$ we find that
\begin{eqnarray}
v(N_a-N_b) &=& 1+\frac{N_t}{4}-\frac{1}{4} N_t \cos(2 \theta) \nonumber \\ 
&-&\frac{1}{2} e^{N_t (-1+\cos(4 \lambda))} N_t \cos(2 \lambda) \sin^2(\theta) \nonumber \\
&-& e^{-2 N_t \sin^2(\lambda)} N_t \sin(2 \lambda) \sin(2 \theta) \, . \label{vanal}
\end{eqnarray}
For $\lambda \ll1$, this simplifies to
\begin{eqnarray}
v(N_a-N_b) &\approx& 1+ \frac{1}{4} \left(N_t+e^{-8 \lambda^2 N_t} N_t (-1+\cos(2 \theta)) \right. \nonumber \\ 
&-& \left. N_t \cos(2 \theta)-8 \lambda e^{-2 \lambda^2 N_t} N_t \sin(2 \theta)\right) .
\end{eqnarray}
\begin{figure}
\includegraphics[width=1.0\columnwidth]{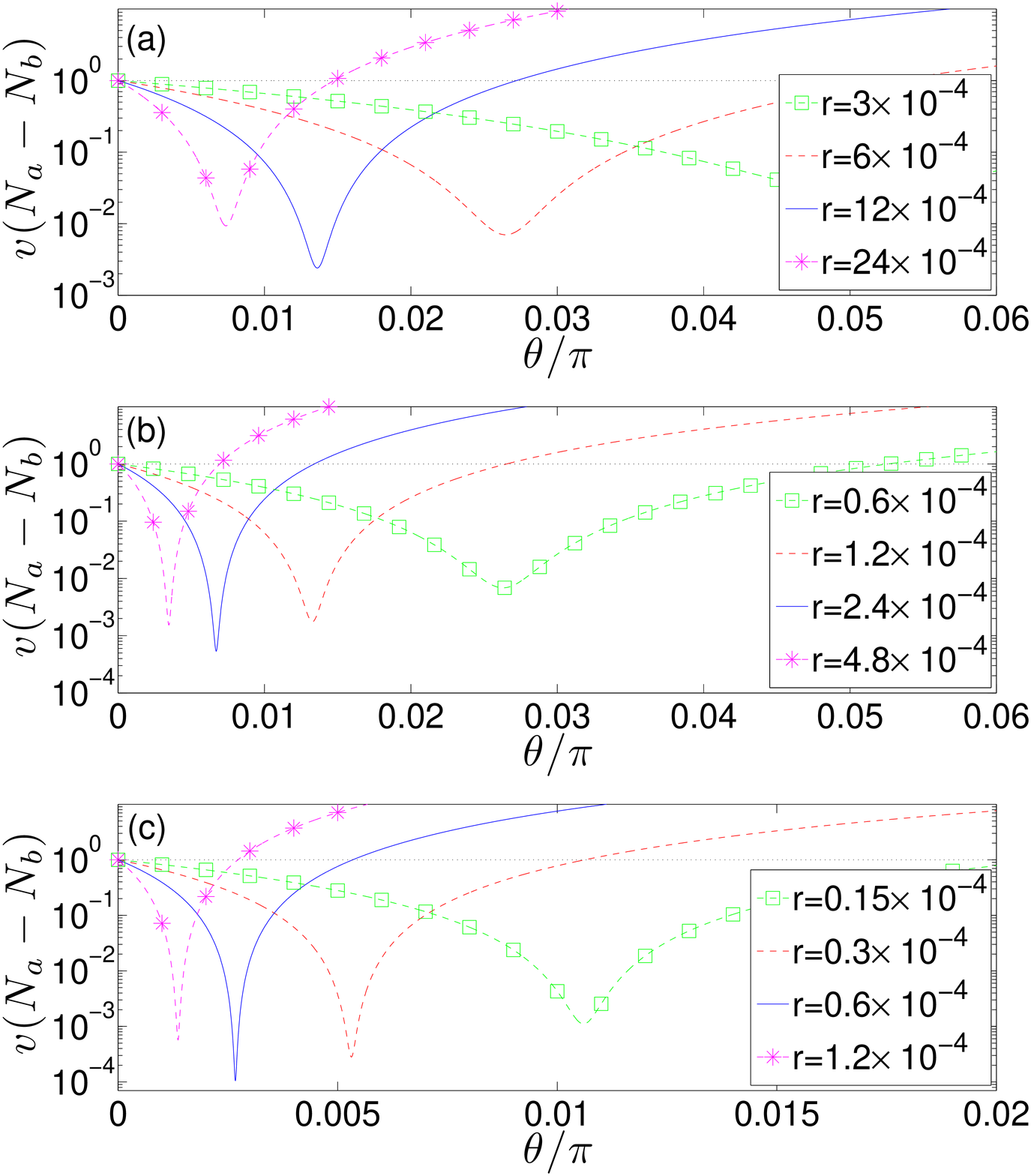}
\caption{\label{fig:vanal} $v(N_a-N_b)$ vs $\theta$ for: (a) $N_t= 10^4$ (b) $N_t=10^5$ (c) $N_t=10^6$.}
\end{figure}
Figure~\ref{fig:vanal} shows \eq{vanal} vs. $\theta$ for several different values of $\lambda$ and $N_t$. As $N_t$ increases, higher levels of squeezing can be obtained. Increasing $\lambda$ beyond a critical amount $\lambda_{\mathrm{opt}}$ begins to degrade the quality of the squeezing. It is better to work  with $\lambda<\lambda_{\mathrm{opt}}$ rather than $\lambda>\lambda_{\mathrm{opt}}$, as the squeezing is more tolerant to slight variations from the optimum value of $\theta$, $\theta_{\mathrm{opt}}$. Figure~\ref{fig:var_vs_r} shows $v(N_a-N_b)$ evaluated at $\theta = \theta_{\mathrm{opt}}$ as a function of $\lambda$ and $N_t$. For large $N_t$, the maximum amount of squeezing approaches
\begin{equation}
v(N_a-N_b)(\lambda_{\mathrm{opt}}, \theta_{opt}) \approx N_t^{-\frac{2}{3}} \label{vopt}
\end{equation}
at
\begin{equation}
 \lambda_{\mathrm{opt}} \approx 0.6 N_t^{-\frac{2}{3}} . \label{ropt}
 \end{equation}
In order to incorporate the spatial dynamics, we calculate the effective squeezing parameter $r$ based on the evolution of the mode functions $u_a(\boldr,t)$ and $u_b(\boldr,t)$ as determined from a GPE simulation. We perform a GPE simulation of the system that was investigated in section \ref{sec2}. This is done by simulating equations (\ref{stoch_eom1}) and (\ref{stoch_eom2}) without the $1/\Delta v$ corrections, and without noise terms in the initial conditions. After obtaining $u_a(\boldr,t)$ and $u_b(\boldr,t)$ we can calculate $\lambda$ from \eq{chi_ij},  \eq{r1_def} and \eq{r2_def}, and obtain $\lambda=7.99\times 10^{-4}$. Figure~\ref{fig:v_vs_vanal} shows a comparison of the squeezing calculated from the semi-analytic model with $\lambda=7.99\times 10^{-4}$ with the full TW result. The optimum squeezing appears at a vastly different value of $\theta$, suggesting that the GPE has drastically overestimated the squeezing parameter. The reason for the large discrepancy is that we have ignored the contribution from the kinetic energy to the phase evolution in \eq{kerrphase1}. Slight differences in the number of particles in each mode cause significant deviations to the spatial dynamics, and hence \eq{kerrphase1} is not a good estimate of the phase evolution of each number state \cite{Li2009}. We note that in some regimes \cite{haine2009} \eq{kerrphase1} \emph{does} give reasonable agreement with the multi-mode TW simulation. However, these situations are when both modes remain close to the ground state of the many-body system. In this paper, the excitations in the system are well beyond the linear regime. 

We will now derive an alternate method to estimate the squeezing parameter, $\lambda$ from the GPE equation. This method is related but not identical to the method used by Li \etal to derive the spin-squeezing dynamics of a multi-mode system \cite{Li2009}. The spin squeezing originates from uncertainty in the number difference coupling to uncertainty in the phase due to the number dependence in the energy of each mode. In the fully quantum simulation, after the first beam splitter, the number difference variance should be $V(N_a-N_b) = N_t$. We can estimate the phase diffusion by calculating the phase from two slightly different GPE simulations, one with an initial beam splitter such that $N_a-N_b = \sqrt{N_t}/2$, the other with $N_a-N_b = -\sqrt{N_t}/2$. That is, two simulations with a difference in $J_z$ equal to the projection noise. Defining the relative phase as
\begin{equation}
\phi_{\mathrm{GPE}} = \mathrm{arg}\left( \int \psi_b^*(2T_\pi, \boldr)\psi_a(2T_\pi, \boldr) \, \dr \right) \, , \label{phasedef}
\end{equation}
our estimate of the phase diffusion relating from this number uncertainty becomes
\begin{equation}
\Delta \phi = \phi_+-\phi_-
\end{equation}
where $\phi_{\pm}$ is the result of evaluating \eq{phasedef} with initial conditions 
\begin{equation}
N_a = \frac{N_t}{2} \pm \frac{\sqrt{N_t}}{4} \, , \quad N_b = \frac{N_t}{2} \mp \frac{\sqrt{N_t}}{4} \, .
\end{equation}
By defining 
\begin{equation}
J_x = \frac{1}{2}\int \psi_b^*(2T_\pi, \boldr)\psi_a(2T_\pi, \boldr) \, \dr \, + c.c. 
\end{equation}
we note that the difference in $J_x$ between the two simulations is approximately
\begin{equation}
\Delta J_x \approx N_t \Delta \phi \, . \label{deltaJ_phi}
\end{equation}

In order to relate this quantity to the squeezing parameter $\lambda$ in the fully quantum two-mode model, we define the $x$ component of the collective spin as
\begin{equation}
J_x = \frac{1}{2}\left(\ahat \bhatd + \bhat \ahatd\right) \, ,
\end{equation}
and note that \eq{kerrstate0} gives
\begin{eqnarray}
V(J_x) &=&  N_t  +\frac{N_t^2}{2}\left( 1  +  \cos 2\lambda (\sinh\left[2 N_t \sin^2 2\lambda \right] \right. \nonumber \\
 &-& \left. \cosh\left[N_t (\cos 4\lambda-1)\right] )\right) \\
&\approx& N_t + 4 \lambda^2 N_t^3 \label{VJ_r} \, .
\end{eqnarray}
for $\lambda\ll1$. By comparing \eq{VJ_r} to the square of \eq{deltaJ_phi}, and noting that for no phase diffusion ($\lambda=0$, $\Delta\phi =0$), \eq{VJ_r} gives $\Delta J_x = \sqrt{N_t}$,  while \eq{deltaJ_phi} gives $\Delta J_x =0$, as it neglects the \emph{zero-point} quantum uncertainty in $J_x$, we obtain
\begin{eqnarray}
\lambda \approx \frac{\Delta \phi}{2\sqrt{N_t}} \, . \label{rphi}
\end{eqnarray}
For the parameters used in Figure~\ref{densmovie}, $\lambda = 9.18\times 10^{-6}$, which is nearly two orders of magnitude less than the value given by \eq{rchi}. Figure~\ref{fig:v_vs_vanal} shows that this gives much better agreement with the 3D TW simulation. The optimum value of $\theta$ is nearly the same, which is an indicator that this is close to the best match with the two-mode model. There is a discrepancy with the maximum level of squeezing obtained, which we attribute to imperfect mode-matching at the final beam-splitter, leading to an overlap of $Q<1$. From \eq{ropt}, we see that $\lambda_{\mathrm{opt}} \approx 2.13 \times 10^{-4}$, indicating that the system is in the regime of being considerably under-squeezed.

\section{Investigation of optimum parameter regime}
As we found in the previous section, although significant squeezing can be obtained via this method, it is a long way from the  maximum allowed by the two-mode model  (\eq{vopt} and \eq{ropt}). We will now investigate how tuning the trapping frequency will effect the degree of squeezing. Tightening the trapping frequency will have three effects. The first is that it will increase the density of the system, which we expect should increase the squeezing rate. The second is that the time taken for the system to perform one `bounce' will be shorter, which will decrease the degree of squeezing, as in this system the time for a `bounce' is always less than the time required for best squeezing. The third effect is that the ratio of kinetic to interaction energy will change, which may cause higher order excitation in our system. In the strongly interacting regime, these excitation frequencies are irrational multiples of each other, so complete spatial re-phasing may not be possible, which will significantly decrease the overlap of the two modes.  Depending on the relative scaling of these competing effects, we may be able to find a regime that gives the maximum amount of squeezing. We found  that a TW simulation assuming spherically symmetry gave excellent agreement with the full, 3D simulation  (see Figure~\ref{fig:v_vs_vanal}), which is convenient, as it uses orders of magnitude fewer computational resources. Specifically, the equations of motion for our complex fields become

\begin{eqnarray} \label{GPE}
i\hbar\frac{\partial \psi_a(r)}{\partial t} &= \mathcal{L}_a \psi_a(r) + \frac{1}{2} \hbar \Omega(t)\psi_b(r) \, , \label{stoch_eom1} \\
i\hbar\frac{\partial \psi_b(r)}{\partial t} &= \mathcal{L}_b \psi_b(r) + \frac{1}{2} \hbar \Omega^*(t) \psi_a(r), \label{stoch_eom2}
\end{eqnarray}
where
\begin{eqnarray}
\mathcal{L}_i &=& \frac{-\hbar^2}{2m}\left(\frac{1}{r^2}\frac{\partial}{\partial r}\left(r^2 \frac{\partial}{\partial r}\right)\right) + \frac{1}{2}m\omega_r^2 r^2 \ \\ 
&+& U_{ii} \left(|\psi_i(r)|^2-\frac{1}{\Delta v}\right) +   U_{ij} \left( |\psi_j(r)|^2 - \frac{1}{2\Delta v} \right), \nonumber
\end{eqnarray}

\begin{figure}
\includegraphics[width=\columnwidth]{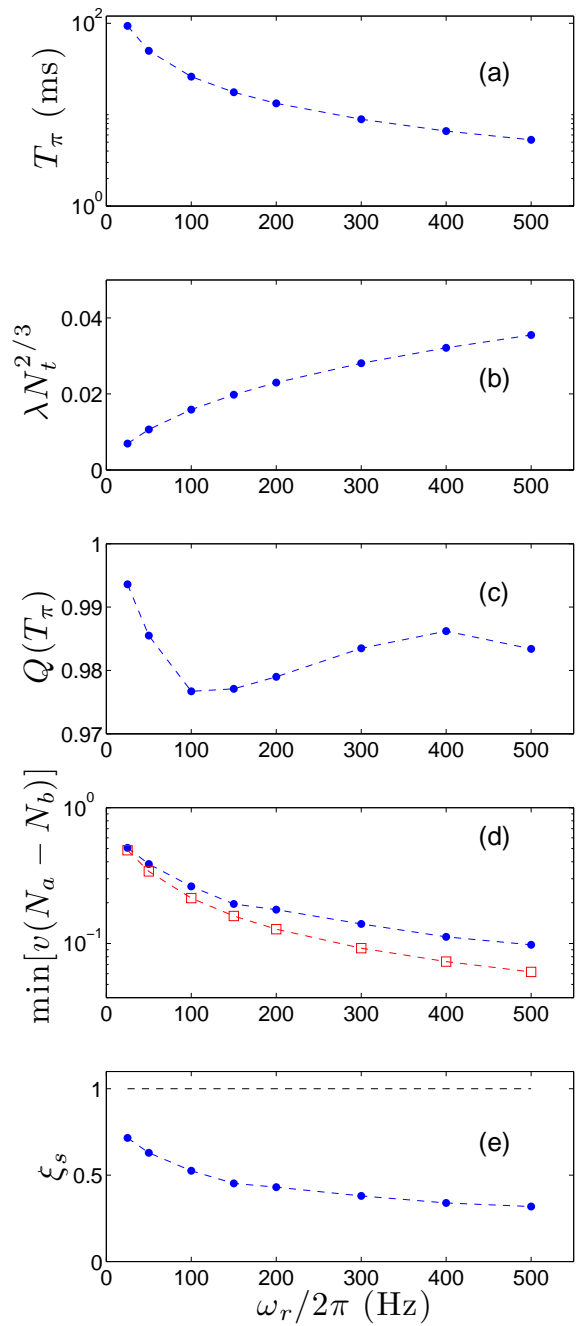}
\caption{\label{paramspam} 1D spherically symmetric TW simulation for different values of $\omega_r$. (a): $T_{\pi}$, the time it takes for one breathing oscillation vs. $\omega_r$. (b): The effective squeezing parameter $\lambda$ as calculated from \eq{rphi} and a 1D spherically symmetric GPE calculation. (c): The overlap $Q$ at the instant of the final beamsplitter. (d): Minimum of $v(N_a-N_b)$ as calculated from a 1D spherically symmetric TW simulation (blue dots), compared to the two-mode analytic result from \eq{vanal}, using the $\lambda$ value from (b). (e) the spin squeezing parameter, $\xi_s$.   }
\end{figure}

Figure~\ref{paramspam} shows the maximum obtainable squeezing for a range of radial trapping frequencies. Increasing $\omega_r$ increases the effective squeezing parameter, $\lambda$, even though $T_{\pi}$ decreases. However, as $\omega_r$ increases, the dependence of $\lambda$ on $\omega_r$ becomes increasingly weak. Even at the maximum value of $\omega_r$ simulated,  $\omega_r=2\pi \times 500$ rad s$^{-1}$, which would be a challenging level of confinement to achieve, $\lambda  \approx 0.035 N_t^{-2/3}$ is approximately a factor of $16$ less than $\lambda_{\mathrm{opt}}$ given by \eq{ropt}, which will give the maximum level of squeezing. As $Q<1$, the actual squeezing is less than the level predicted by the two-mode model.  As the level of squeezing increases, a slight imperfection in mode-matching has a larger detrimental effect for the squeezing, which is why the discrepancy between the TW and two-mode models increases with $\omega_r$. 

\begin{figure}
\includegraphics[width=\columnwidth]{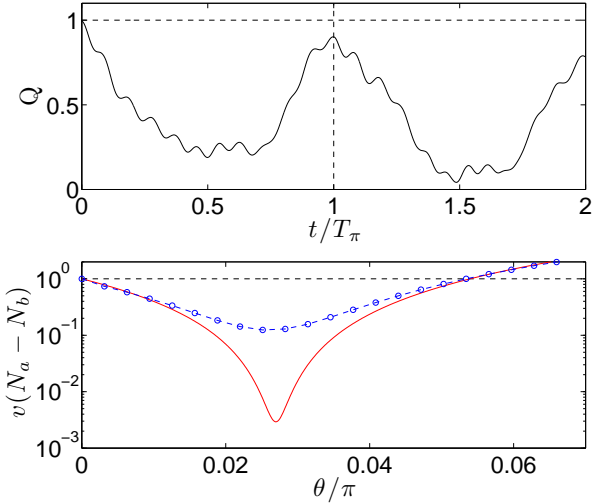}
\caption{\label{cylfig} 3D TW simulation for $\omega_x=\omega_y = 2\pi\times 500$ rad s$^{-1}$, $\omega_z = 2\pi\times 100$ rad s$^{-1}$. Top: $Q$ vs. $t$. Bottom: $v(N_a-N_b)$ vs. $\theta$ calculated from the 3D TW simulation (blue circles), and \eq{vanal} (red solid trace), using $\lambda$ calculated from a 3D GPE  simulation and \eq{rphi}.} 
\end{figure}

Counter-intuitively, we can increase $\lambda$ by \emph{decreasing} the trapping frequency in one dimension, while keeping the same confinement in the other two directions. This is because $T_{\pi}$ increases, due to the breathing mode in the weaker trapping direction, while the density remains high due to confinement in two tightly confined directions. Figure~\ref{cylfig} shows $Q$ and $v(N_a-N_b)$ for $\omega_x=\omega_y = 2\pi\times 500$ rad s$^{-1}$, $\omega_z = 2\pi\times 100$ rad s$^{-1}$. In this parameter regime, the system undergoes complicated nonlinear evolution, and $T_\pi$ increases to $56.4\,$ms. However, the complicated evolution causes the revival in $Q$ to be much less than for the spherically symmetric case. This is partly due to breathing oscillations occurring at vastly different frequencies in the different directions, but also due to exchange of energy between the breathing mode of the tight directions ($x$ and $y$) with higher order modes in the weak ($z$) direction. This evolution has the desired effect in increasing $\lambda$ to $6.16\times 10^{-5}$ (up from $1.26\times 10^{-5}$ for the spherically symmetric $\omega_r = 2\pi\times 500$ rad s$^{-1}$ case). However, as the overlap is vastly decreased, most of this increased squeezing is lost when considering the multi-mode TW simulation, and it performs worse than the spherically symmetric case. 

\begin{figure}
\includegraphics[width=\columnwidth]{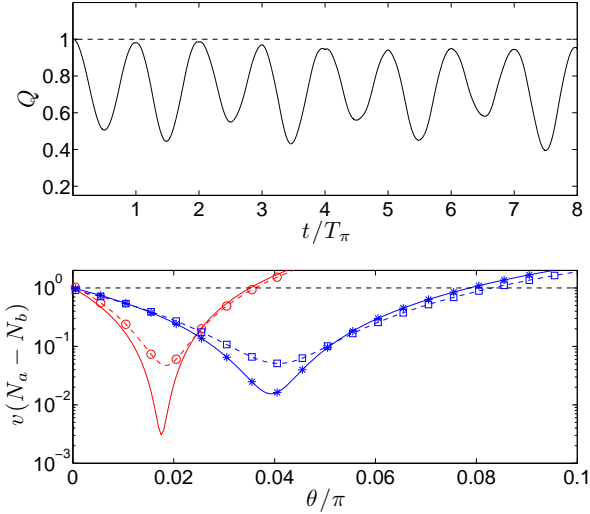}
\caption{\label{doublebounce} Spherically symmetric TW simulation of multiple $\pi$ pulses. Top: The visibility $Q$ as a function of time for a sequence of several $\pi$ pulses separated by free-evolution time. After the initial $\pi/2$ pulse at $t=0$, the $\pi$ pulses are repeated with a period of $T_\pi$. Bottom: $v(N_a-N_b)$ calculated from the spherically symmetric TW simulation for the double bounce (blue squares) and quadruple bounce (red circles) schemes. $v(N_a-N_b)$ is also calculated from \eq{vanal} (blue stars: double bounce, red solid trace: quadrupole bounce), using $\lambda$ calculated from a spherically symmetric GPE and \eq{rphi}. }
\end{figure}

In an attempt to increase the level of squeezing, we try multiple iterations of the scheme, that is, repeating the sequence of $\pi$ pulses and free evolution periods multiple times before the final beamsplitter in order to increase the interaction time and presumably increase $\lambda$. Figure~\ref{doublebounce} shows $v(N_a-N_b)$ and $Q$ for a total of $2$ and $4$ times as much total free evolution time, for $\omega_r = 2\pi\times 500$ rad s$^{-1}$. The free evolution time between each $\pi$ pulse was always kept fixed at $T_\pi = 5.3$ ms. We will refer to these two schemes as ``double bounce'' and ``quadruple bounce'' respectively.  While there are still quasi-periodic revivals in the visibility, there is a slight decay as the number of iterations is increased. The increase in $\lambda$ is a factor of approximately $2$ and $4$ for the double and quadruple bounce schemes respectively, which is still a factor of $7$ and $3.5$ less than $\lambda_{\mathrm{opt}}$. Due to the nonlinear dependence of \eq{vanal} on $\lambda$, this leads to a reduction in $v(N_a-N_b)$ of  approximately $4$ and $20$ respectively. However, when considering the full multi-mode TW dynamics, the decrease in visibility degrades the level of squeezing, and their is only a factor of  $\sim 2$ improvement for the double bounce scheme, and minimal further improvement for the quadruple bounce scheme. As increasing the evolution time is likely to exacerbate other detrimental effects, such as increased particle loss or decoherence due to technical noise, it is unlikely that it will be advantageous to consider multiple bounces. 

\section{Relation to precision metrology and spin squeezing.}\label{sec4}
We have demonstrated how to use multi-mode dynamics to enhance the one-axis twisting rate in order to prepare a state with reduced fluctuations in particle number difference. In order to use this state for interferometry with sensitivity beyond the standard quantum limit (SQL), the output from the final beam-splitter of the one-axis twisting scheme would be used as the input to a two-port Mach-Zehnder interferometric scheme. That is, a 50/50 beamsplitter, followed by a relative phase shift $\phi$ between components $a$ and $b$ caused by the physical process one wishes to examine, followed by a final 50/50 beamsplitter. The coupling operations occur on a timescale much faster than the motional dynamics, so the motional dynamics can be neglected during the beamsplitter phases. Furthermore, we will assume that the time between the beamsplitters, $t_{\mathrm{hold}}$ is short compared to the timescale for motional dynamics. Typically the sensitivity of atom interferometry scales linearly with $t_\mathrm{hold}$, so it may be desirable to increase $t_\mathrm{hold}$ beyond the regime of validity of this approximation. We will discuss the implications of this below. Using these approximations, we can solve for the dynamics analytically in the Heisenberg picture
\begin{eqnarray}
\psihat_a(\boldr, t_{\mathrm{out}}) &=& -i\psihat_a(\boldr, t_{\mathrm{in}}) \sin\left(\frac{\phi}{2}\right) \nonumber \\ 
&-&  i\psihat_b(\boldr, t_{\mathrm{in}}) \cos\left(\frac{\phi}{2}\right) \label{MZ1} \\
\psihat_b(\boldr, t_{\mathrm{out}}) &=& -i \psihat_a(\boldr, t_{\mathrm{in}}) \cos\left(\frac{\phi}{2}\right) \nonumber \\
&+&  i\psihat_b(\boldr, t_{\mathrm{in}}) \sin\left(\frac{\phi}{2}\right) \label{MZ2}
\end{eqnarray}
where $\psihat_{a,b}(\boldr,t_{\mathrm{in}})$ is the field operator after the final beam splitter of the squeezing sequence (and input of the Mach-Zehnder interferometer), and $\psihat_{a,b}(\boldr, t_{\mathrm{out}})$ is the field operator after the final beamsplitter of the Mach-Zehnder interferometer. At this point ($t=t_{\mathrm{out}}$), the number difference is measured, from which we can estimate the value of the applied phase shift. 
The phase sensitivity of the device is given by
\begin{equation}
\Delta \phi = \frac{\sqrt{ V\left(N_a(t_{\mathrm{out}})-N_b(t_{\mathrm{out}})\right)}}{\abs{\frac{d}{d \phi}\langle \left(N_a(t_{\mathrm{out}})-N_b(t_{\mathrm{out}})\right)\rangle}} \, .
\end{equation}
For uncorrelated input states, we recover the standard quantum limit $\Delta \phi = 1 / \sqrt{N_t}$ \cite{dowling}. Noticing that $\hat{N}_a-\hat{N}_b = 2\hat{J}_z$ and using \eq{MZ1} and \eq{MZ2}, we find $\hat{J}_z(t_{\mathrm{out}}) = \sin \phi \, \hat{J}_x(t_{\mathrm{in}}) - \cos\phi \, \hat{J}_z(t_{\mathrm{in}})$. If we ensure that our input state lies along the $J_x$ axis (that is, $\langle \hat{J}_z(t_{\mathrm{in}})\rangle = \langle \hat{J}_y(t_{\mathrm{in}})\rangle =0$, which can always be achieved by suitable choice of a deterministic phase shift), the slope of our signal will be maximum at $\phi = 0$ (or $\pi$). In this case, we can write the maximum phase sensitivity as 
\begin{equation}
\Delta \phi = \frac{\sqrt{\langle \hat{J}_z^2(t_{\mathrm{in}})\rangle}}{\langle \hat{J}_x(t_{\mathrm{in}}) \rangle}  = \frac{\xi_s}{\sqrt{N_t}}  \, ,
\end{equation} 
where $\xi_s \equiv \sqrt{N_t \langle \hat{J}^2_z\rangle}/\langle \hat{J}_x\rangle$ is the usual spin squeezing parameter \cite{wineland94,sorensen2001}. \fig{paramspam} shows $\xi_s$ calculated immediately after the final beam-splitter for different values of $\omega_r$ from the spherically symmetric TW simulation. For $\omega_r = 2\pi \times 500$ rad s$^{-1}$, $\xi \approx 0.32$, which indicates an interferometric phase uncertainty $\sim 3$ times better than for uncorrelated particles, or equivalent to using $9$ times as many uncorrelated particles.

In writing \eq{MZ1} and \eq{MZ2} we have neglected the motional dynamics of each component during the Mach-Zehnder process. This is equivalent to assuming that the overlap between the modes is unchanged during the interferometer process (however, in calculating $\xi_s$ we have taken into account the effect of imperfect overlap at the input to the Mach-Zehnder). To gain any significant benefit from the spin squeezing, the interferometry scheme must involve a high-degree of overlap between the two modes. In a trapped configuration, this would limit the duration $t_{hold}$ of the interferometer to very short times, before the multi-mode dynamics from the strong nonlinear interactions begin to degrade the overlap. Alternatively, setting $t_{hold}$ to multiples of $T_{\pi}$ would also achieve high-overlap due to the revivals in $Q$. For some applications, such as inertial sensing, an atom interferometer that operates in free fall is desirable, as it is isolated from vibrational noise (aside from that coupled in through the control lasers). After the relative number squeezing is created, the clouds could be expanded by releasing, or adiabatically expanding, the confining potential. For an inertial sensor, momentum separation between the two modes is required, which could be achieved accelerating one of the modes with a state-selective Bragg transition or Bloch oscillation after the wave packets are sufficiently dilute \cite{macdonaldET2013}. Expanding the BEC has the added benefit of reducing the density, which will reduce any deleterious effects due to nonlinear interactions, such as phase-diffusion. 

\section{Summary}\label{sec5}
We have shown that spatial dynamics can be used to enhance the rate of one-axis twisting, to produce significant spin-squeezing without the use of a Feshbach resonance or state-dependent dynamic potentials in atoms such as $^{87}$Rb where the squeezing rate would otherwise be too low. We find that generally tighter traps are better, leading to higher squeezing, which is achieved much more quickly, which will be important in the presence of loss processes such as collision with background gas. Using a cylindrically symmetric potential causes the effective squeezing parameter to increase, but the time taken to achieve squeezing is also increased, and the increased dynamical excitations limit the degree of squeezing achievable. Performing multiple bounces seems promising, but this also eventually causes a loss of overlap due to multi-mode excitations. We found that the best achievable squeezing for $1.5\times 10^5$ atoms is $v(N_a-N_b)\approx 0.047$, with an overlap $Q=0.95$, by performing a four-bounce sequence in the tightest trap we considered, $\omega_r = 2\pi\times 500$ rad s$^{-1}$. It seems unlikely that this scheme could yield significantly higher squeezing, as the achievable squeezing is very sensitive to the degree of overlap. However, even though this is considerably less than the theoretically achievable limit predicted by \eq{vanal}, we are considering a large number of atoms, which will yield a large absolute increase in sensitivity for an interferometric device, equivalent to an increase of a factor of $19$ in the atom number. One of the benefits of incorporating the multi-mode excitations into the squeezing scheme, rather than trying to remove them altogether, is that it opens the way for one-axis twisting experiments with larger samples of atoms where previous schemes have been limited in the number of particles to try and maintain single-mode dynamical behaviour.

\section{Acknowledgements}
We would like to acknowledge useful discussions with Matthew Davis, Joel Corney, Jacopo Sabatini, Tod Wright, Chao Feng, and Michael Hush. This work was supported by the Australian Research Council Discovery Project No. DE130100575. \\


\begin{thebibliography}{99}
\bibitem{chu99} A. Peters, K. Y. Chung, and S. Chu, Nature {\bf 400}, 849 (1999).
\bibitem{chu2001} A. Peters, K. Y. Chung, and S. Chu, Meteroliga, {\bf 38}, 25, (2001).
\bibitem{kasevich02} J. M. McGuirk, G. T. Foster, J. B. Fixler, M. J. Snadden, and M. A. Kasevich, Phys. Rev. A {\bf 65}, 033608 (2002).
\bibitem{kasevich97} T. L. Gustavson, P. Bouyer, and M. A. Kasevich, Phys. Rev. Lett. {\bf 78}, 2046 (1997).
\bibitem{kasevichG} J. B. Fixler, G. T. Foster, J. M. McGuirk, and M. A. Kasevich, Science {\bf 315}, 5808 (2007).
\bibitem{altinET2013} P. A. Altin, M. T. Johnsson, V. Negnevitsky, G. R. Dennis, R. P. Anderson, J. E. Debs, S. S. Szigeti, K. S. Hardman, S. Bennetts, G. D. McDonald, D. Pulford, L. D. Turner, J. D. Close and N. P. Robins, New J. Phys. {\bf 15}, 023009 (2013).
\bibitem{biraben_alpha} R. Bouchendira, P. Clade, S. Guellati-Khelifa, F. Nez, F. Biraben, Phys. Rev. Lett. {\bf 106}, 080801 (2011).
\bibitem{kasevich_GW} P. W. Graham, J. M. Hogan, M. A. Kasevich, and S. Rajendran, Phys. Rev. Lett. {\bf 110}, 171102 (2013). 
\bibitem{debs2011} J. E. Debs, P. A. Altin, T. H. Barter, D. D\"{o}ring, G. R. Dennis, G. McDonald, R. P. Anderson, J. D. Close, and N. P. Robins,  Phys. Rev. A {\bf 84}, 033610 (2011). 
\bibitem{szigeti2012} S. S. Szigeti, J. E. Debs, J. J. Hope, N. P. Robins, J. D. Close, New Journal of Physics {\bf 14}, 023009 (2012). 
\bibitem{dowling} J. P. Dowling, Phys. Rev. A {\bf 57}, 4736 (1998).
\bibitem{oberthaler2010} C. Gross, T. Zibold, E. Nicklas, J. Esteve, and M. K. Oberthaler, Nature {\bf 464}, 1165 (2010).
\bibitem{treutlein2010} M. F. Riedel, P. Bohl, Y. Li, T. W. Hansch, A. Sinatra, and P. Treutlein Nature {\bf 464}, 1170 (2010).
\bibitem{lucke2011} B. L\"{u}cke, M. Scheer, J. Kruse, L. Pezze, F. Deuretzbacher, P. Hyluss, O. Topic, J. Peise, W. Ertmer, J. Arlt, L. Santos, A. Smerzi, and C. Klempt, Science {\bf 334}, 773 (2011). 
\bibitem{leroux2010} I. D. Leroux, M. H. Schleier-Smith, V. Vuletic, Phys. Rev. Lett. {\bf 104}, 073602 (2010).
\bibitem{ueda} M. Kitagawa and M. Ueda, Phys. Rev. A {\bf 47}, 5138 (1993).
\bibitem{sorensen2001} A. Sorensen, L. -M. Duan, J. I. Cirac, P. Zoller, Nature {\bf 409}, 63 (2001).
\bibitem{sorensen2002} A. S. Sorensen, Phys. Rev. A {\bf 65} 043610 (2002).
\bibitem{gross2012} C. Gross, J. Phys. B {\bf 45}, 103001, (2012).
\bibitem{mertes2007} K. M. Mertes, J. W.Merrill, R. Carretero-Gonz\'alez, D. J. Frantzeskakis,  P. G. Kevrekidis, and D. S. Hall, Phys. Lett. {\bf 99}, 190402 (2007).
\bibitem{steel} M. J. Steel, M. K. Olsen, L. I. Plimak, P. D. Drummond, S. M. Tan, M. J. Collett, D. F. Walls, and R. Graham, Phys. Rev. A {\bf 58} 4824 (1998).
\bibitem{sinatra2002} A. Sinatra, C. Lobo, Y. Castin, J. Phys. B {\bf 35}, 3599 (2002).
\bibitem{johnsson2007} M. T. Johnsson and S. A. Haine, Phys. Rev. Lett.  {\bf 99} 010401 (2007).
\bibitem{haine2009} S. A. Haine and M. T. Johnsson, Phys. Rev. A {\bf 80}, 023611 (2009). 
\bibitem{haine2011} S. A. Haine and A. J. Ferris, Phys. Rev. A {\bf 84} 043624 (2011).
\bibitem{Egorov2012} B. Opanchuk, M. Egorov, S. Hoffmann, A. I. Sidorov and P. D. Drummond, EPL {\bf 97}, 50003 (2012).
\bibitem{dall2009} R. G. Dall, L. J. Byron, A. G. Truscott, G. R. Dennis, M. T. Johnsson, and J. J. Hope, Phys. Rev. A {\bf 79}, 011601(R) (2009).
\bibitem{dennis2010} G. R. Dennis and M. T. Johnsson, Phys. Rev. A {\bf 82} 033615 (2010).
\bibitem{johnsson2013} M. T. Johnsson, G. R. Dennis and J. J. Hope, New J. Phys. {\bf 15} 123024 (2013). 
\bibitem{haine2013} S. A. Haine, Phys. Rev. Lett. {\bf 110}, 053002 (2013).
\bibitem{Li2008} Y. Li, Y. Castin, A. Sinatra, Phys. Rev. Lett. {\bf 100}, 210401 (2008).
\bibitem{sinatrareview2012} A. Sinatra, J. C. Dornstetter, Y. Castin, Front. Phys {\bf 7}, 86 (2012).
\bibitem{ferrini2011} G. Ferrini, D. Spehner, A. Minguzzi, and F. W. J. Hekking, Phys. Rev. A {\bf 84} 043628 (2011).
\bibitem{anderson2009} R. P. Anderson, C. Ticknor, A. I. Sidorov, and B. V. Hall, Phys. Rev. A {\bf 80}, 023603 (2009). 
\bibitem{egorov2011} M. Egorov, R. P Anderson, V. Ivannikov, B. Opanchuk, P. Drummond, B. V. Hall, and A. I. Sidorov, Phys. Rev. A {\bf 84}, 021605 (2011).
\bibitem{atlin} P. A. Altin, G. McDonald, D. D\"{o}ring, J. E. Debs, T. H. Barter, J. D. Close, N. P. Robins, S. A. Haine, T. M. Hanna, and R. P. Anderson, New Journal of Physics, {\bf 13} 065020 (2011).
\bibitem{blackie2008} P. B. Blakie, A.S. Bradley, M.J. Davis, R.J. Ballagh, and C.W. Gardiner, Advances in Physics, {\bf 57}, 363 (2008).
\bibitem{mtj2007} M. T. Johnsson and J. J. Hope, Phys. Rev. {\bf A 75}, 043619 (2007).
\bibitem{walls} D. F. Walls and G. J. Milburn, Quantum Optics (Springer, Berlin, 1994).
\bibitem{olsenwig} M. K. Olsen, A. S. Bradley, Optics Comm. {\bf 282}, 3924 (2009). 
\bibitem{xmds} G. R. Dennis, J. J. Hope, and M. T. Johnsson, Comput. Phys. Commun. {\bf 184}, 201-208 (2013). 
\bibitem{wineland94} D. J. Wineland, J. J. Bollinger, W. M. Itano, and D. J. Heinzen, Phys. Rev. A {\bf 50}, 67 (1994).  
\bibitem{ho1996} T. L. Ho, V. B. Shenoy, Phys. Rev. Lett {\bf 77}, 3276 (1996).
\bibitem{Li2009} Y. Li, P. Treutlein, J. Reichel, and A. Sinatra, Eur. Phys. J. B {\bf 68}, 365 (2009).
\bibitem{macdonaldET2013} G. D. McDonald, C. C. N. Kuhn, S. Bennetts, J. E. Debs, K. S. Hardman, M. T. Johnsson, J. D. Close, N. P. Robins, Phys. Rev. A {\bf 88}, 053620 (2013).
\end{thebibliography}
\end{document}